# 3D Printing Autoclavable PPE on Low-Cost Consumer 3D Printers

**Authors:** *Hannelore Hemminger[a] and *Xiaoyu (Rayne) Zheng[ab]*

**Affiliations:**

[a]*Department of Civil and Environmental Engineering, & Department of Mechanical and Aerospace Engineering, University of California, Los Angeles, CA 90095, USA*
[b]*Department of Mechanical Engineering, Virginia Tech, Blacksburg, VA 24061, USA*

**\*Corresponding author:**

Email address: rayne@seas.ucla.edu (X. Zheng)

**Abstract:**

During the COVID-19 pandemic, medical facilities began using 3D printed PPE sourced from their own print labs, makerspaces, universities, and individuals with 3D printers to fill the gaps in supply as traditional manufacturing was not widely enough distributed nor quick enough to scale to widespread spikes in demand. However, to date this PPE has been limited to low-temperature, easy to print thermoplastics which are not compatible with autoclave sterilization and must be sterilized by hand washing methods. Herein, we present a method for 3D printing a temperature resistant nylon copolymer on a common low-cost 3D printer. We show that the resulting parts can be autoclaved without deformation, and conduct uniaxial tensile testing showing that autoclaving the material does not result in substantial degradation of material properties. As a result, we demonstrate the capability to manufacture autoclavable PPE on low-cost consumer 3D printers with only minor modification.

1.  Introduction

The onset of a global pandemic has revealed a need for widely distributed manufacturing of medical Personal Protective Equipment (PPE). With demand for PPE spiking and traditional manufacturing unable to scale production quickly and flexibly enough to match it, medical facilities around the world began using 3D printed PPE sourced from their own 3D printers, local makerspaces and universities, or even individuals who own one or more 3D printers [1]. Previously, work to bring a distributed supply chain to autoclavable PPE has focused on creating new designs for 3D printers capable of printing high-temperature thermoplastics, such as the open-source Cerberus 3D printer from Michigan Tech [2]. There has also been some analysis done on design efficacy of individual PPE designs printed on commercially available printers, but little to no research has taken place on expanding the printing materials used by these commercial printers, which presently is typically Polylactic Acid (PLA), Polyethylene Terephthalate Glycol (PETG), Acrylonitrile Butadiene Styrene (ABS), Acrylonitrile Styrene Acrylate (ASA), or nylon [3] [4].

While easy to print, PLA has a Vicat softening temperature of around 62 °C [5]. This means it cannot be sterilized by steam autoclave, which takes place at 121 °C, and must be sterilized with other methods such as surface disinfectants to avoid severe warping and drooping of the part. While larger facilities such as hospitals are likely to have other compatible automated sterilization methods such as low-temperature hydrogen peroxide gas plasma [6], smaller or older facilities without such modern equipment available currently need to manually disinfect 3D printed PPE, which can take medical personnel several minutes per item [7]. Steam autoclave is a disinfecting process which uses pressurized steam at 121 °C to sterilize objects by way of heat [8] [9]. It is an effective and automated sterilization process and allows for objects to be sterilized in large quantities. 3D printing PPE in an autoclavable material would save valuable time for medical professionals on the front lines of the pandemic, as well as building confidence that the parts are fully disinfected, since sterilizing 3D printed parts with surface disinfectants can be inconsistent due to the large number of small gaps and crevices which result from the 3D printing process. To allow steam autoclave to be used as a disinfecting method for 3D printed PPE, it is necessary to expand the materials used to print them to include those which can survive autoclave temperatures without major deformation. Since consumer 3D printers are not capable of high extrusion temperatures and do not have high temperature heated chambers [10] [11] [12] [13], they are incapable of producing parts made from high-temperature engineering thermoplastics that extrude at well over 300 °C, such as Polyether Ether Ketone (PEEK) or Polyetherimide (PEI), which would typically be chosen when autoclavability is required. To save the time of medical professionals, it is therefore desirable to demonstrate the use of an autoclavable thermoplastic with widely available consumer 3D printers with as little modification as possible to provide widespread, autoclavable PPE manufacturing capabilities.

In this paper, we demonstrate the 3D printing of a nylon copolymer (CoPA) manufactured by Polymaker and test its resistance to autoclave conditions by printing and autoclaving test components which consist of large bridges and overhangs to test for unacceptable deformations [14]. Tensile testing specimens with dimensions according to ASTM D638 Type V are also 3D printed and tested on a tensile testing machine (Instron) to check if and how mechanical properties of the material are degraded by autoclave. The print quality of the CoPA material is raised to acceptable levels by designing a very simple heated chamber, capable of heating to up to 50 °C without causing damage to the printer, consisting of a cardboard and clear plastic sheet enclosure built around the printer with a commonly available 500-watt personal space heater and temperature controller available from Amazon.

## 2. Theory and Hypothesis

Common engineering thermoplastics with high temperature resistance also have a high glass transition temperature, with PEEK and PEI having glass transition temperatures of 143 °C and 217 °C respectively [15]. As a thermoplastic cools, once it reaches its glass transition temperature it can no longer release internal stresses, which in 3D printing causes warping of the printed part due to thermal contraction of the polymer [16]. Large differences between a 3D printing polymer's glass transition temperature and the build chamber temperature can therefore cause a large amount of warping, and thermoplastics with a high glass transition temperature, including PEEK and PEI, typically require a high temperature heated build chamber to print successfully. Other factors in part warping include a polymer's thermal expansion coefficient and crystalline behavior, although these behaviors can be more complicated and are frequently not reported by 3D printing filament manufacturers. Some thermoplastics – including PEEK, but not PEI, maintain physical strength at temperatures above their glass transition temperature [15]. One measure of what temperature a polymer softens at is its Vicat softening temperature, making the Vicat softening temperature a simple method of predicting whether a polymer will survive autoclave cycles without deforming [15]. Therefore, to find a 3D printing thermoplastic which will survive autoclave but be printable on a Prusa MK3S or similar machine, the authors hypothesize that a polymer should have a Vicat softening temperature well above the 121 °C of steam autoclave, a low glass transition temperature, and a recommended printing temperature safely under the 300 °C limit of the Prusa MK3S hotend. Semi-crystalline polymers should be of particular interest due to their different modulus characteristics in the temperature region near their glass transition temperature compared to amorphous plastics [17]. A plot of Vicat softening temperature (VST) vs. glass transition temperature (Tg) of many 3D printing materials is shown below in **Figure 1** [15] [18] [19] [20] [21] [22] [23].

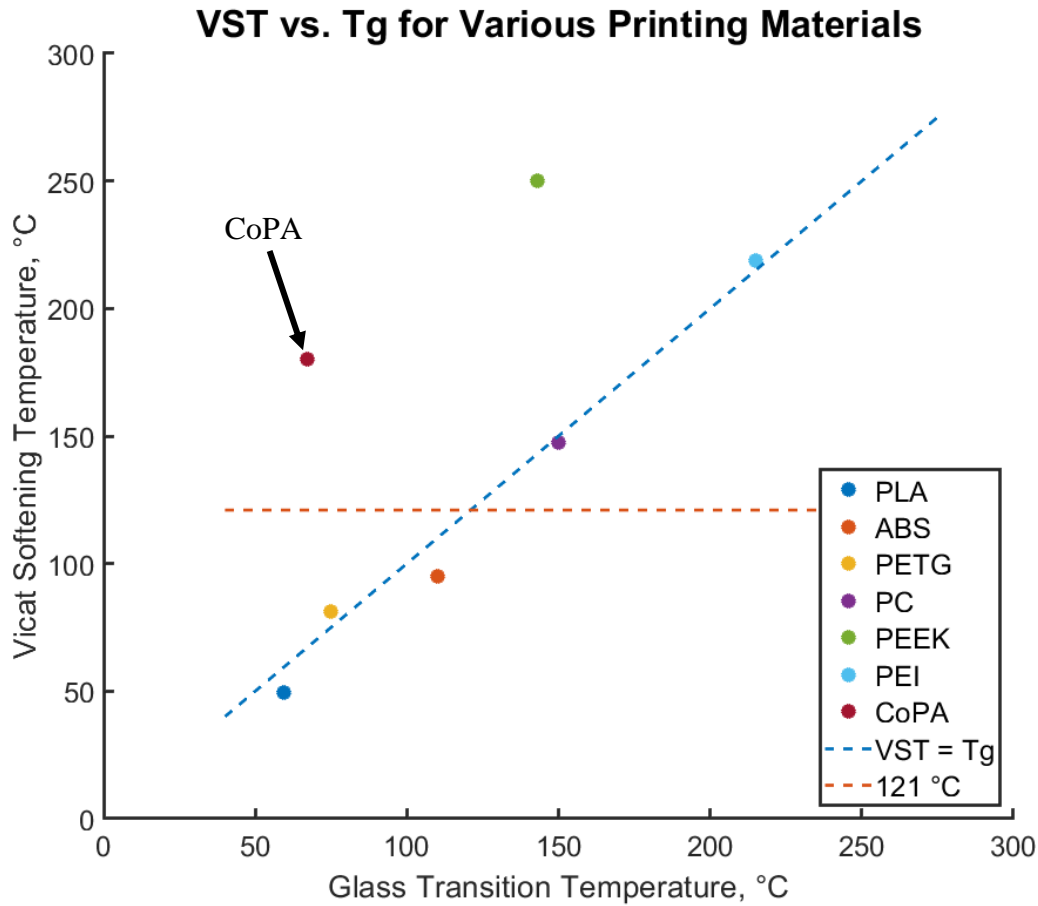

*Figure 1:* *Vicat softening vs. glass transition temperatures of many 3D printing filament types and blends. Notice that most polymers follow the line VST = $T_g$, but that notable outliers include CoPA and PEEK. *For PEEK, the continuous service temperature is shown in place of Vicat softening temperature due to incomplete data.*

CoPA, a nylon-based copolymer made by Polymaker, has a Vicat softening temperature of 180 °C, a glass transition temperature of 67 °C, and a recommended printing temperature of 250-270 °C, making it a good candidate for testing [23]. The authors hypothesize that it should be possible to print CoPA on a Prusa MK3S or any similar printer, since the manufacturer's recommended build environment temperature is only 40-60 °C [23], and a chamber temperature of up to 45-50 °C should be tolerable for the motors, electronics, and 3D printed components of most consumer machines.

3. Materials and Methods

**Consumer 3D Printer selection**

The Prusa MK3S 3D printer was selected to represent a commonly available consumer 3D printer. Available at $800 USD and with a production volume of over 60,000 printers in 2019 [24], it is popular in makerspaces and universities as well as with individual owners, making it widely available for distributed manufacturing of PPE. It has also already been used in other studies on 3D printing PPE [3]. With a maximum extrusion temperature of 300 °C, it cannot extrude high temperature engineering thermoplastics such as PEEK and PEI, but is capable of extruding CoPA which prints at 250 - 270 °C.

**Heated enclosure setup**

The heated enclosure was constructed from cardboard, tape, a space heater and temperature controller, and clear sheet plastic to provide a viewing window, along with two twist ties to help secure the enclosure door. A 5-sided oversized box was constructed around the printer, with the 5$^{th}$ (front) side of the box hollowed out to an approximately 3-inch frame and copied a second time. The center of the 2$^{nd}$ copy was filled with the clear plastic sheeting to create a see-through door, which was hinged to the front top edge of the enclosure. Finally, a hole was pierced in the top for the temperature probe and a cutout was made in the back of the right panel to install the space heater. The completed enclosure with all components installed is shown in **Figure 2**.

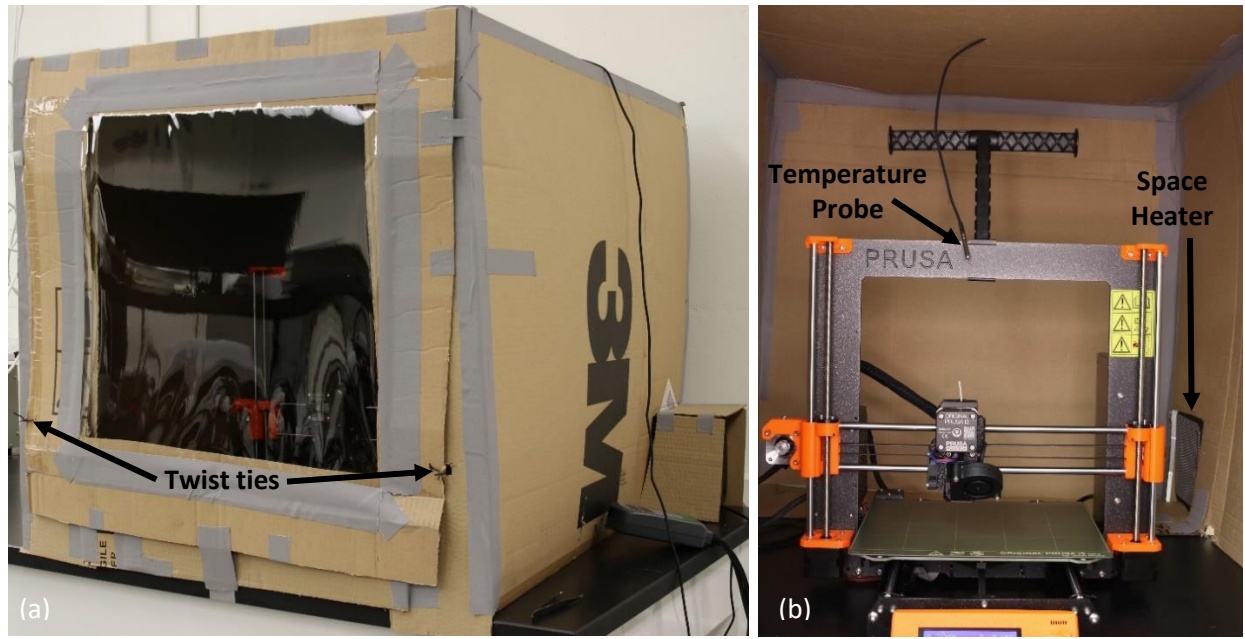

*Figure 2:* *Cardboard heated build chamber. (a): Closed chamber. Note the use of twist ties to secure the chamber door and clear plastic sheet used to provide visualization of the print. (b): Inside of the chamber. Note the position of the space heater (AmazonBasics) and the temperature probe of the off-the-shelf temperature controller (Inkbird ITC-308).*

While a much higher quality enclosure could be constructed using standard aluminum extrusions and plastic panels for example, the authors' intent is to demonstrate that such engineering time and materials are unnecessary to create a satisfactory heated chamber which functions well enough to raise the print quality of the CoPA material to an acceptable level. The described enclosure can be constructed in only an hour or two with minimal tools, and only requires materials and tools which are commonly available and are likely already owned by most individuals and makerspaces.

**3D Printing of CoPA**

To initially test for drooping and structural integrity in autoclaved components, test pieces including a tower with 4 increasingly large overhangs and a bridge were printed and subjected to a 121 °C steam autoclave sterilization cycle. Test pieces were then inspected for any substantial deformation and compared to test pieces printed in PLA and Acrylonitrile Styrene Acrylate (ASA). To check for unacceptable degradation in mechanical properties, 20 ASTM D638 type V tensile testing specimens, with a cross-section of 3.2x3.2mm, were manufactured using perimeter-only 100% infill, and half were autoclaved. In each group, 5 specimens were dried prior to testing while 5 were tested while acclimated to atmospheric moisture content. Prior to printing, all filament of each material was dried thoroughly to avoid quality issues caused by atmospheric moisture absorption. During testing, it was noticed that prolonged heat soaks at drying temperatures (several days or more over many overnight drying cycles) were causing the filament to gradually degrade and become brittle and fragile. For this paper, authors used vacuum sealing filament containers with desiccant packs to store and dry the filament to avoid heating it repeatedly, but recommend that others make an airtight filament dry-box with Bowden fittings to allow for printing from the dry-box directly. The manufacturer states that CoPA can be abrasive to brass nozzles when used extensively [14], although no nozzle abrasion was noted by the authors during the printing of approximately 1.5 kg of CoPA through an uncoated brass nozzle.

## 4. Results and Discussion

**Part fidelity and deformation tests**

In initial testing, no heated chamber was used, and the print quality with CoPA was consistently poor. Overhangs showed a tendency to curl upward when cooling to room temperature and the print quality also degraded as the filament absorbed atmospheric moisture, since CoPA is a hygroscopic material. The authors suspect that the crystalline behavior of CoPA, which has a crystallization temperature of 128 °C, is likely responsible for the upward warping of the overhangs, due to the observation during printing that this particular category of warping occurred while the deposited plastic was still well above its glass transition temperature. After constructing the heated chamber and drying the filament thoroughly before printing, print quality improved substantially, and with further tuning of print settings it was possible to achieve good print quality even for medium to high complexity objects such as the popular Benchy boat model [25]. The final major print settings used for CoPA (Polymaker), set in Prusa Slicer, are shown below in Table 1 alongside the print settings used for the deformation test objects for PLA (Polymaker) and ASA (Polymaker).

*Table 1: Print Settings*

| Print Material | Extrusion Temperature, °C | Bed Temperature, °C | Enclosure Temperature, °C | Layer Height, mm | Extrusion Width, mm |
|---|---|---|---|---|---|
| PLA | 210 | 60 | 20 | 0.2 | 0.4 |
| ASA | 260 | 105 | 20 | 0.2 | 0.4 |
| CoPA | 260 | 70 | 48 | 0.2 | 0.4 |

For large objects, especially objects with sharp corners, a large brim size was used for CoPA prints to improve build platform adhesion, or custom adhesion pads were added to the part design at sharp corners contacting the build platform.

The results of the deformation testing are shown in **Figure 3**. After autoclave, parts printed from both PLA and ASA exhibited clear thermal failure with severe warping and drooping, while parts printed from CoPA held their shape.

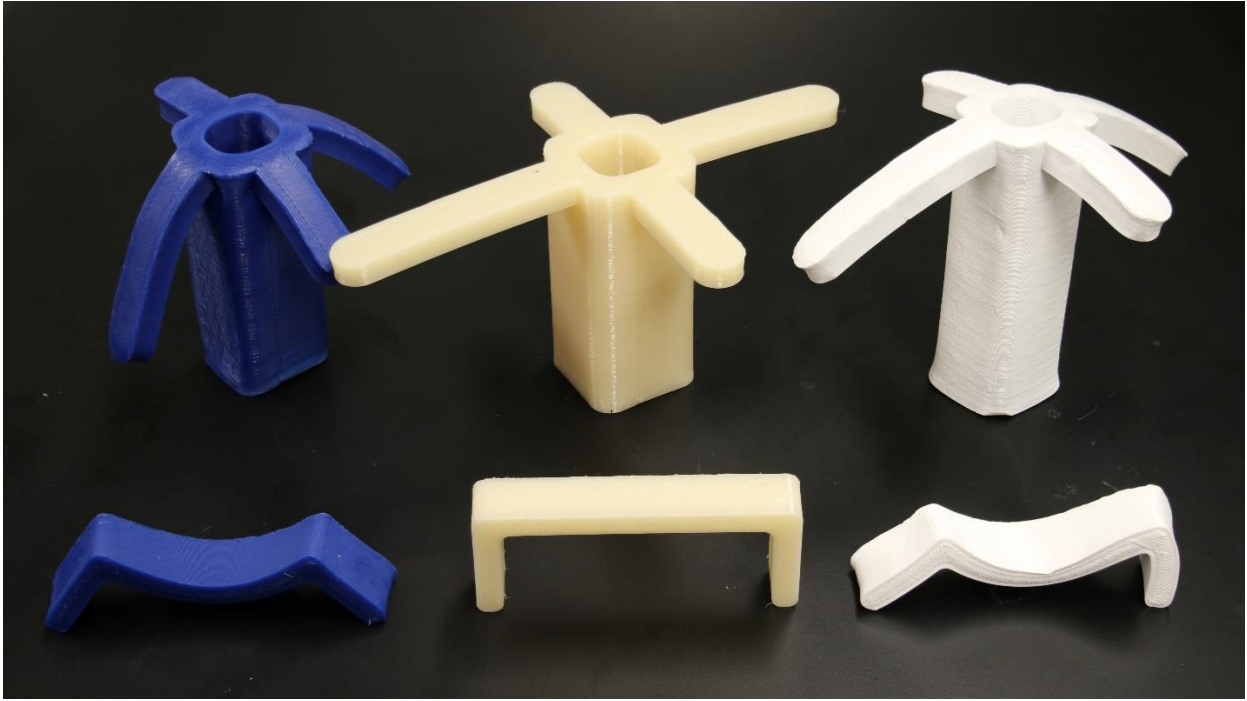

*Figure 3:* Post-autoclaving deformation testing results. Blue: PLA, Beige: CoPA, White: ASA. Top: Testing of cantilevers of various lengths. Bottom: Testing of bridges. PLA and ASA both fail to withstand autoclave temperatures for all cantilevers as well as the bridge, while CoPA has no detectable deformation on the bridge experiment and only minimal deformation on even the largest cantilevers tested.

**Design and Printing for Uniaxial Tensile Test**

Initially, tensile testing was difficult due to specimen failure at a weak point inherent in the 3D printed part where the extrudates spread outward as the specimen widens. The point where a central filling extrudate joined the spreading extrudates had lower strength than the rest of the specimen, and consistently failed first before any necking occurred in the specimen as shown below in Figure 3. Unsuccessful solution attempts included varying the infill pattern, re-melting the weak point area with a soldering iron, and packing salt around the prints and reflowing them in an oven. The weak point was eventually eliminated by processing the 3D model as 5 different pieces, separated by 1 micron, consisting of a central spar running the length of the specimen and

4 reinforcing wings which form the remaining width of the specimen grip. When using perimeter only infill by setting the number of perimeters to be large as well as setting the part combination distance to be small compared to the piece separation, this results in a printed specimen with all 5 pieces fused with a consistent central spar with no inherent weak points running the full length of the specimen. Every extrudate in the testing area of the specimen runs parallel to the direction of tension, yielding accurate results of the material strength. A comparison of the toolpaths and resulting tested specimens is shown below in **Figure 4** before and after manipulating the toolpath.

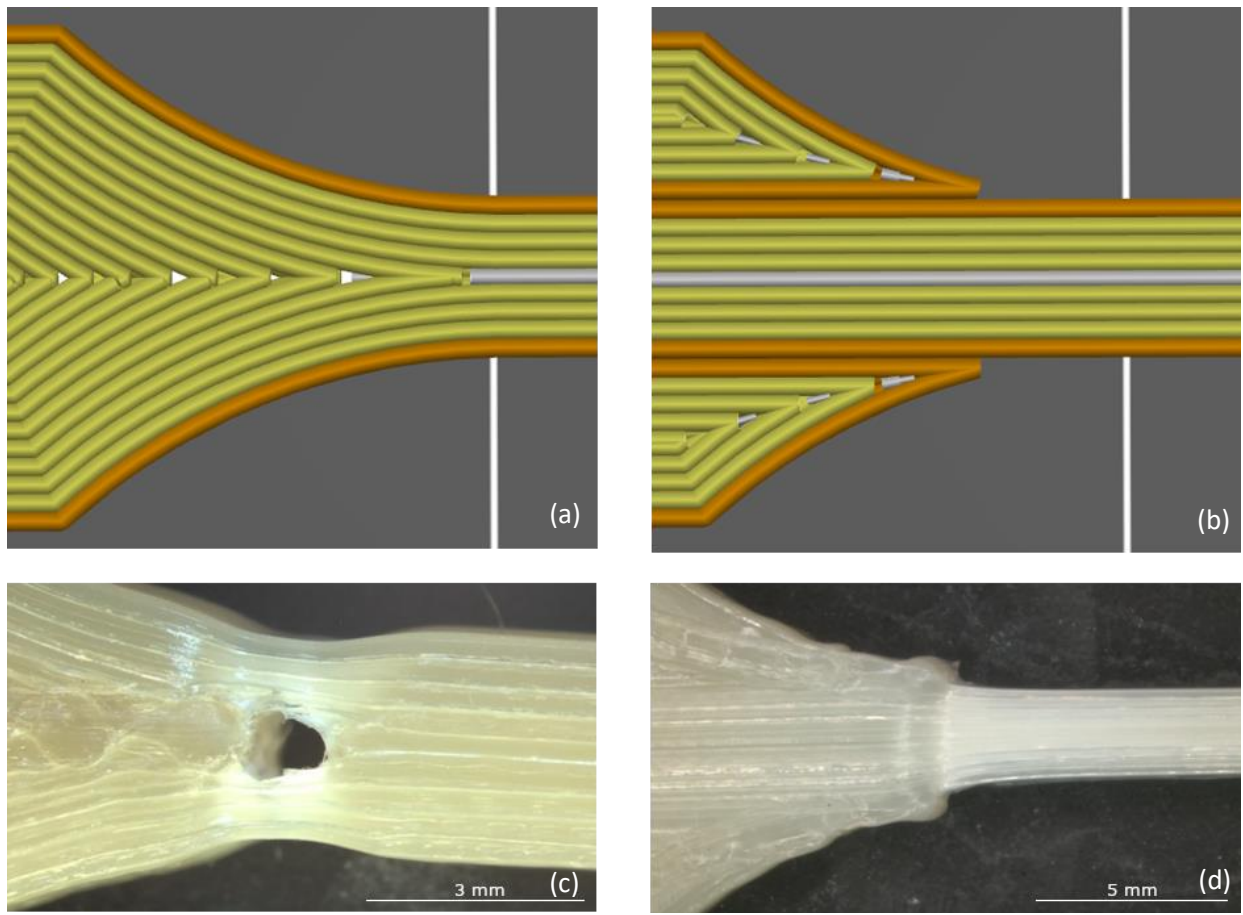

*Figure 4: Toolpaths and tested specimens before and after manipulating the printer toolpath via part splitting and slight separation. (a) and (c): Before toolpath manipulation. (b) and (d): After toolpath manipulation via part splitting and separation.*

## Uniaxial Tensile Properties of 3D printed CoPA

The results of the tensile testing for mechanical degradation are shown in Table 2, reporting the mean engineering values of Ultimate Tensile Strength (UTS) and mechanical strain at UTS. Stress-strain curves showing the first 50% strain are also shown for qualitative comparison in **Figure 5**.

*Table 2: Tensile testing results*

|  | Mean UTS, MPa | UTS Standard Deviation, MPa | Mean Strain at UTS, % | Strain at UTS Standard Deviation, % |
| --- | --- | --- | --- | --- |
| Non-Autoclaved, Wet | 69.4 | 1.9 | 16.6 | 0.31 |
| Autoclaved, Wet | 65.8 | 0.7 | 17.1 | 0.75 |
| Non-Autoclaved, Dry | 73.4 | 0.9 | 17.8 | 0.32 |
| Autoclaved, Dry | 64.4 | 2.1 | 17.3 | 0.26 |

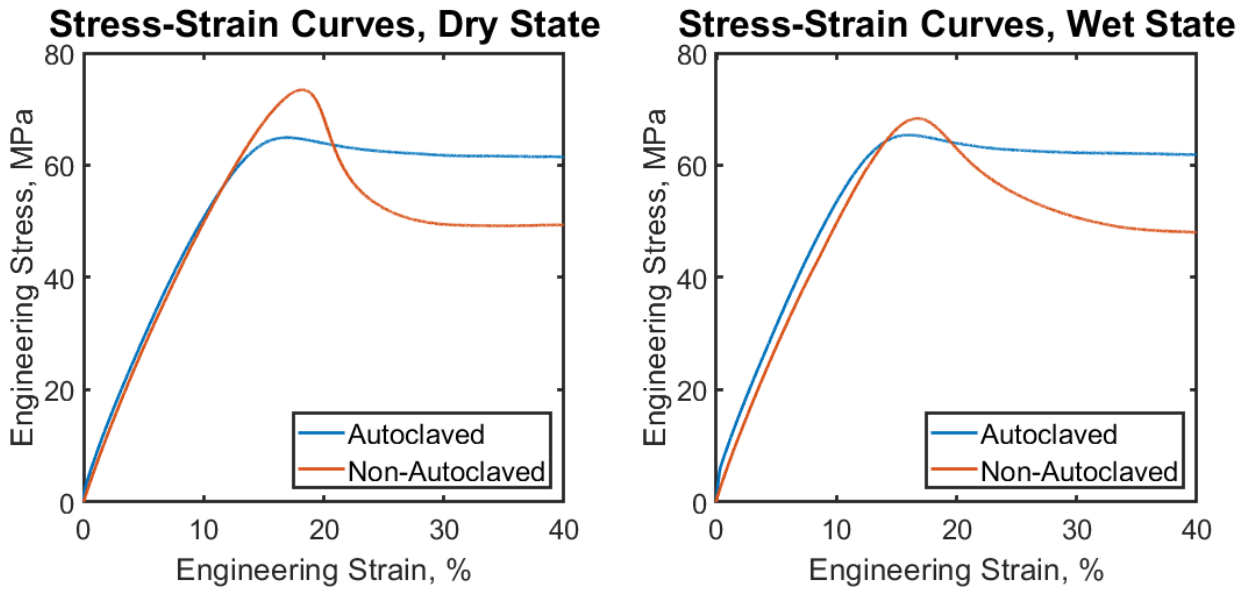

***Figure 5:** Representative tensile testing engineering stress-strain curves. Left: Samples after being dried in a sealed container with desiccant at reduced pressure. Right: Samples allowed to reach equilibrium with atmospheric moisture ('wet state'). Since specimens are 3D printed, there was variation between individual stress-strain curves, so the curves presented are the closest to a representative sample among the samples tested.*

The table shows that the material withstands autoclave while maintaining strength and with little change to the engineering strain at Ultimate Tensile Strength (UTS). There are several qualitative features of interest in the stress-strain curves shown above. Most notably, the autoclaved specimens on average have a slightly lower UTS, particularly for the dry specimens, but a substantially larger stress during the necking region of the curve. This increased necking stress is likely the result of annealing taking place during autoclave resulting in lower residual internal stresses and an increased resistance to slip in its semi-crystalline structure. The typical difference between dry and wet samples is also much lower for the autoclaved specimens, which suggests that the method of drying (2+ weeks in low pressure with desiccant packs) may be insufficient for removing the large amount of moisture collected during the autoclave process. Therefore, the authors recommend that future work use a vacuum oven to further reduce the pressure and lightly heat specimens to dry them more thoroughly if it is desired to study the dry state of the material. However, since the properties of the wet material lend itself well to face shield frames and other PPE as they are somewhat more flexible and therefore more comfortable, the authors do not recommend any drying of face shield frames after autoclaving and suggest that the frames go directly back into use after autoclave. It is also possible that the annealing caused by the autoclave temperatures somehow reduces the material's sensitivity to moisture content. More study of the basic material properties with respect to annealed and non-annealed states is suggested for future work. More study is also recommended on the number of times the material can be autoclaved without substantial degradation.

A face shield frame was printed with CoPA and subjected to a sterilization autoclave cycle. Only a small amount of deformation was noticed, and the part remained fully operational after the autoclave cycle, demonstrating that the material is suitable for manufacturing 3D printed autoclavable PPE. A face shield frame printed with CoPA is shown below in **Figure 6**.

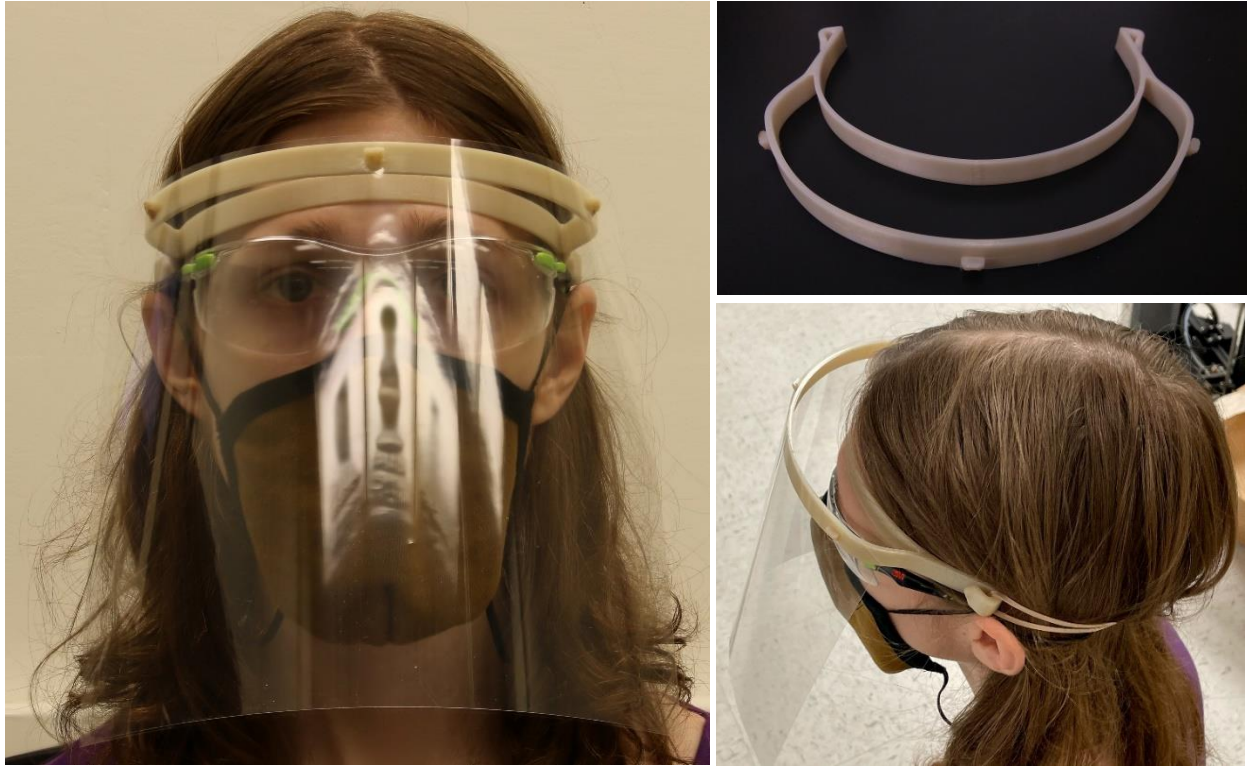

*Figure 6: A researcher wearing a face shield 3D printed from CoPA. After autoclaving face shields printed from CoPA there is only a small amount of deformation, and the part remains fully usable.*

## 5. Conclusion

We presented a new material for 3D printing autoclavable personal protective equipment. We show that, unlike other commonly used materials such as PLA and ABS, CoPA, with its high Vicat softening temperature, can withstand autoclave temperatures without visible deformation and retains its mechanical properties post testing. Unlike other high temperature polymers such as PEEK and PEI, which require industrial level 3D printers not typically accessible to mass consumers, CoPA with its relatively lower glass transition temperature can be printed with acceptable quality with only simple modifications to a commonly available consumer level 3D printer via a heated build chamber. Additionally, our uniaxial tensile experiments reveal that not only are ultimate tensile strength and elastic limit not substantially degraded when autoclaved, some properties, including some post-necking hardening, are also moderately improved, likely attributable to the annealing effect during the autoclave process. We demonstrated 3D printing of face shields that remain functional after being subjected to autoclaving. This work points to a viable path to use widely available consumer 3D printers for autoclavable PPE manufacturing.

**Acknowledgements**

X. Zheng and H. Hemminger would like to thank Dr. Shaily Mahendra and Yu (Rain) Miao for their autoclaving assistance and Huachen Cui and Zhenpeng Xu for their technical support in testing initial print settings. The authors would like to thank 3M Young Faculty Gift Award and Startup support from UCLA.

**Conflicts of Interest**

The authors declare no conflicts of interest.

## 6. Appendices

### A. Enclosure Construction

The heated enclosure was constructed by simply cutting out panels of cardboard to construct an oversized box around the printer, measuring 26 x 32 x 26 inches to allow plenty of clearance room for the filament roll and convenience of operation. All cardboard for the walls and ceiling of the box was simply roughly cut with a pair of scissors, and the box construction was accomplished with the use of regular gaffer's tape, although any similar tape or suitable adhesive may be substituted. To create a door frame on the front, a cut out was made in the front of the box with an Exacto knife, leaving approximately 3-4 inches on all sides. For the door itself, another matching window was cut out of another piece of cardboard using an Exacto knife and a 10-gauge clear vinyl sheet was taped in its place. Once again, any suitable clear plastic may be substituted if the materials used are stable at the intended chamber temperature. If no clear sheet material is available, the authors suggest creating a small viewport instead of cutting out a large window and covering it on both sides with clear tape. The assembled door was then hinged to the top of the enclosure with more tape and the other sides left free to allow opening and closing the door. To secure the door for printing, 2 sets of 2 holes were pierced through the door and doorframe in the bottom left and bottom right corners, allowing the door to be secured by passing a twist-tie through the holes and twisting it closed. The cardboard door was found to have a slight tendency to warp due to the temperature gradient between the inside and outside of the chamber as well as tightening of the plastic sheet. To reduce air loss and improve chamber temperature stability, more tape can be used as needed to secure the door. Alternatively, any convenient rigid rods may be used to form and attach a rigid support structure around the outside of the door to prevent the warp, such as simple wooden dowels. To accommodate the space heater, the heater's profile was traced to the cardboard and cut out from the lower back corner of one side panel, allowing the heater to be placed with the output face only slightly into the enclosure with the rest of the heater outside of it. If a better seal is required to reach desired chamber temperatures, cardboard flaps can be taped to the outside of the side panel to get a tighter fit after heater placement. Finally, the temperature probe was added by piercing a hole in the top of the enclosure through which the probe was lowered until the tip was at the same height as the top of the printer frame. The completed enclosure with all components installed is shown in **Figure 2**.